\input{aipcheck}

\documentclass[
    ,final            % use final for the camera ready runs
  ]
  {aipproc}

\layoutstyle{6x9}

\def \D0 {D\O }

\begin{document}

\title{Precision Calibration of the \D0 \ HCAL in Run II}

\classification{01.30.Cc, 07.20.Fw, 29.40.Vj}

\keywords      {calorimeter, calibration, hadronic, {D\O } detector}

\author{ Kriszti\'an Peters}{
  address={School of Physics \& Astronomy, University of Manchester,\\
Manchester M13 9PL, UK\\ \vspace{.3cm} {\rm for the \D0 \ Calorimeter Algorithm Group}}
}

%\author{<author2>}{
%  address={<common address for author2 and author3>}
%}

%\author{<author3>}{
%  address={<common address for author2 and author3>}
%  ,altaddress={<author1 address>} % additional visiting address
%}

\begin{abstract}
Most of the physics analyses at a hadron collider rely on a precise
measurement of the energy of jets in the final state. This requires a
precise {\em in situ} calibration of the calorimeter with the final detector
setup. We present the basic procedure and results of the {\em in situ}
gain calibration of the D0 HCAL in Run II. The gain calibration works
on top of the pulser-based calibration of the readout electronics and
is based entirely on collision data. 

%and the
%measurement of the calorimeter response to single hadrons. In addition
%to calibration, the single hadron response can be further used to
%improve the jet energy measurements. It is an essential input to
%energy flow algorithms, jet energy scale determination for heavy quark
%jets and as a feedback to the detector simulation development.
\end{abstract}

\maketitle

%%%%%%%%%%%%%%%%%%%%%%%%%%%%%%%%%%%%%%%%%%%%
%% MAINMATTER
%%%%%%%%%%%%%%%%%%%%%%%%%%%%%%%%%%%%%%%%%%%%

\section{Introduction}

The detailed description of the \D0 \ Calorimeter can be found in
Run~I and Run~II instrumentation papers \cite{D0nimRunI, D0nimRunII}.
Let us briefly summarize some of its basic aspects.

The \D0 \ calorimeter is segmented into towers in $\eta$ and
$\phi$. The precision towers divide the calorimeter in 64 segments in
the $\phi$ and 72 segments in the $\eta$-direction. Each precision
tower consists of four physical layers in the EM part and four or more
physical layers in the hadronic part. The calorimeter is also divided
into trigger towers, which are 2x2 arrays of precision towers,
dividing the calorimeter into 32 segments in $\phi$, and 37 in
$\eta$. Trigger towers are the smallest calorimeter units seen by the
Level 1 trigger.

\begin{figure}[t]%\hspace{-1.5cm}
\includegraphics[width=.8\textwidth]{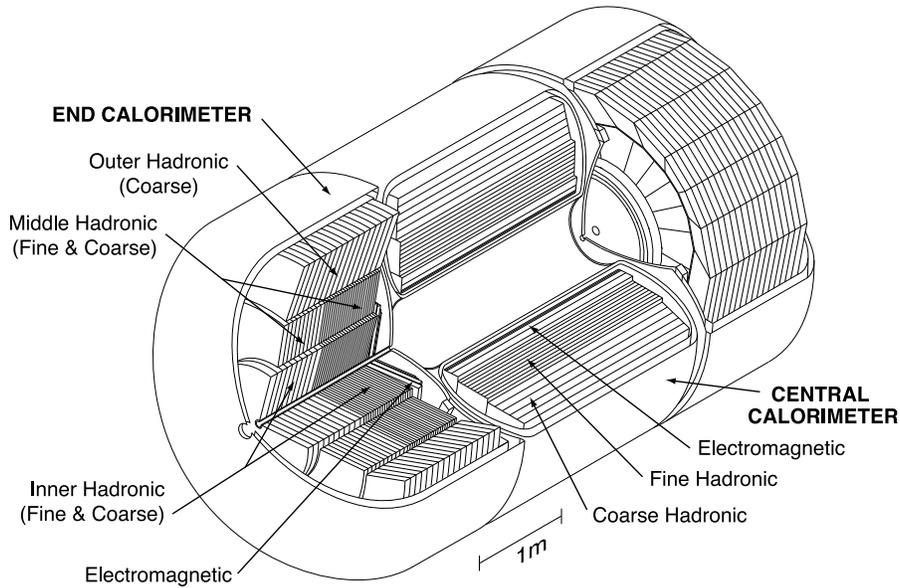}
\caption{Isometric view of the \D0 \ central and two end calorimeters in Run II.} 
\label{calo} 
\end{figure}

We distinguish a central section covering pseudorapidities
$|\eta|$ up to $\approx 1.1$, and two end calorimeters (EC) that
extend coverage to $|\eta|\approx 4.2$, with all three housed in
separate cryostats, Fig.\ref{calo}. These are sampling LAr
calorimeters with mainly Uranium absorber plates.  A calorimeter basic
unit consists of a 3~mm thick plate of an absorber material, 2.3~mm
liquid Argon gap and a signal board consisting of a Copper pad
surrounded by G10 insulator coated with high resistivity epoxy. The
absorber is grounded and the pad is kept at positive voltage
of~2000~V. Charge induced at the pads gives the physical signal. The
electron drift time across the 2.3~mm gap is approximately 450~ns.

Although for Run~II the calorimeter itself is unchanged from Run~I,
the charge integration time has been reduced from $\sim$2.2~$\mu$s in
Run~I to $\sim$260~ns in Run~II, resulting in an enhanced sensitivity
to the finite mechanical precision of the calorimeter.  Additional
mechanical boundaries like module edges, any non-uniformities in the
LAr gap and Uranium plate widths or possible board bendings are
directly related to the amount of charge collected and the response of
the modules and cells. In adittion, the associated readout electronics have been
largely redesigned to address the need for a shorter shaping and
readout time and the need for analog buffering to store the data until
a Level 1 Trigger decision becomes available.

The Run~II upgrade strongly influenced the calorimeter in an indirect
way as well. The amount of dead material in front of the calorimeter
increased significantly with the upgrade and it is non-uniformly
distributed. This material comes from pre-shower detectors, the
solenoid, fiber tracker and silicon vertex detector. Together with the
cryostat walls the particles traverse at least of 3.7 $X_0$
before they reach the calorimeter. The amount of dead material depends
also significantly on the angle of incidence of the measured particles
and increases with the pseudo-rapidity.

Due to this significant changes for the upgrade, the calorimeter had
to be calibrated again. Moreover, it was essential to obtain a
calorimeter Run~II calibration {\em in situ} with the final detector
setup.

\section{Calibration procedure}

The calibration procedure for the \D0 \ calorimeter contains two
parts: calibration of the readout electronics using pulser data, and
correction of non-uniformities due to mechanical variations in the
detector using collision data.

The basic idea of the electronics calibration is to send a pulse of
known charge into the readout, and to compare it to the measured
charge. In this way we identify technical problems in the electronics,
e.g. dead channels and correct for the channel-by-channel differences
in the response. Pulses of different heights are used to probe the
full dynamic range of every readout channel. In this way, the response
of every single channel can be linearized, and the gains of the
different channels can be equalized.

The gain calibration of the \D0 \ calorimeter factorize into two
parts: the calibration of the EM calorimeter and the calibration of
the hadronic calorimeter. For this two parts of the calorimeter we
determine the energy scale (i.e. a multiplicative correction factor),
if possible per cell. Both parts of the calorimeter have been
calibrated in two steps. First, the $\phi$-intercalibration to reduce
the number of degrees of freedom, where special triggered Run II data
was used. Second, the $\eta$-intercalibration to get access to the
remaining degrees of freedom, as well as the absolute scale of the EM
calorimeter. For the $\eta$-intercalibration we used $Z\to e^+e^-$
events for the EM calorimeter and QCD dijet events for the hadronic
calorimeter.

The best standard candle for the absolute calibration of the EM
calorimeter is the $Z$-peak, which is well known from LEP measurements.
%and serves also as an absolute scale of the EM calorimeter. 
However, we lack of statistics to use the $Z$-peak alone in
calibrating on a tower or cell level. Therefore we used special
triggered EM data to intercalibrate over rings of fixed $\eta$. Once
the $\phi$-degree of freedom is eliminated, the amount of $Z$ events
is sufficiently high to absolutely calibrate each intercalibrated
$\eta$-ring.

For this purpose the reconstructed $Z$ mass is written in terms of the
electron energies and their opening angle. The electron energies are
evaluated as the raw energy measurement from the calorimeter plus a
parametrized energy-loss correction from a detailed detector
simulation. Calibration constants are multiplicative to the raw
cluster energy of each cell. A set of calibration constants is then
determined that minimize the experimental resolution on the $Z$ mass
and gives the correct LEP measured value. After the fully calibrated
EM calorimeter, we address the calibration of the hadronic part.

\section{$\phi$-intercalibration of the \D0 \ HCAL}

Due to the fact that the $p\bar p$ beams in the Tevatron are
unpolarized, the energy flow in the direction transverse to the beam
should not have any azimuthal dependence. Based on this, we can use
an energy flow method with the following basic principle:

Consider in each case a given $\eta$-bin of the calorimeter. Measure
the density of calorimeter objects above a given $E_T$ threshold as a
function of $\phi$. With a perfect detector this density would be flat
in $\phi$. Assuming that any $\phi$-non-uniformities are due to energy
scale variations, the uniformity of the detector can be improved by
applying multiplicative calibration factors to the energies of the
calorimeter objects in each $\phi$-region in such a way that the
candidate density becomes flat in $\phi$. 

A special trigger was designed to record efficiently data for the
hadronic $\phi$-intercalibration. It requires a transverse energy
threshold of 5 GeV in the Trigger Tower at Level~1, then it requires
at Level~2 that 5 GeV is in the hadronic part of the tower and finally
it tightens the hadronic transverse energy cut for a Precision Tower
at Level~3 to 7 GeV. Data for the $\phi$-intercalibration was
taken during normal physics running.  The quality of the recorded data
was studied in detail to separate failures in redout electronics from
gain miscalibrations.  Systematic uncertainties from trigger
non-unifomities were avoided by placing trigger and offline cuts
sufficiently above the trigger conditions.

The task of deriving $\phi$-intercalibration constants
for cells in towers at given $\eta$ is divided
into the two following steps:
\begin{itemize} 
  \item Finding a tower calibration constant, which is a
  multiplicative factor for all cells in the tower, such that tower
  occupancies above an $E_T$ threshold are equalized in $\phi$.
\item Fitting layer calibration constants, which then intercalibrate cells
  within the tower. This is performed using cell energy fraction
  distribution shapes which are compared to the $\phi$-averaged
  reference shape.
\end{itemize}
Since the above two steps influence each other, the procedure of layer
and tower calibration has to be iterated until stability is
reached. The final $\phi$-intercalibration constants are the products
of these layer and tower constants. 

\begin{figure}[t]
  \includegraphics[width=.5\textwidth,bbllx=26pt,bblly=252pt,bburx=284pt,bbury=443pt,clip=,]{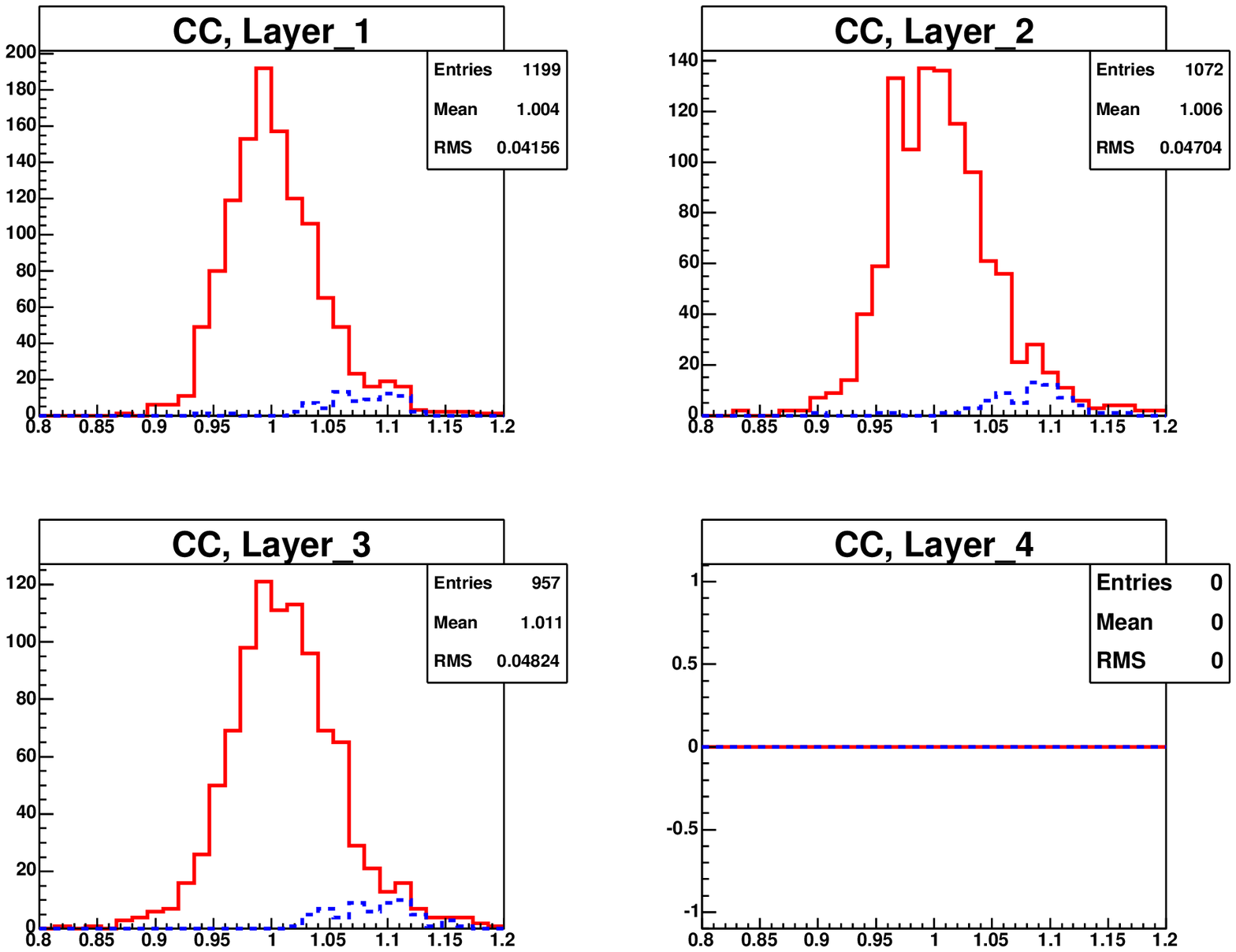}
  \includegraphics[width=.49\textwidth,bbllx=26pt,bblly=260pt,bburx=284pt,bbury=458pt,clip=,]{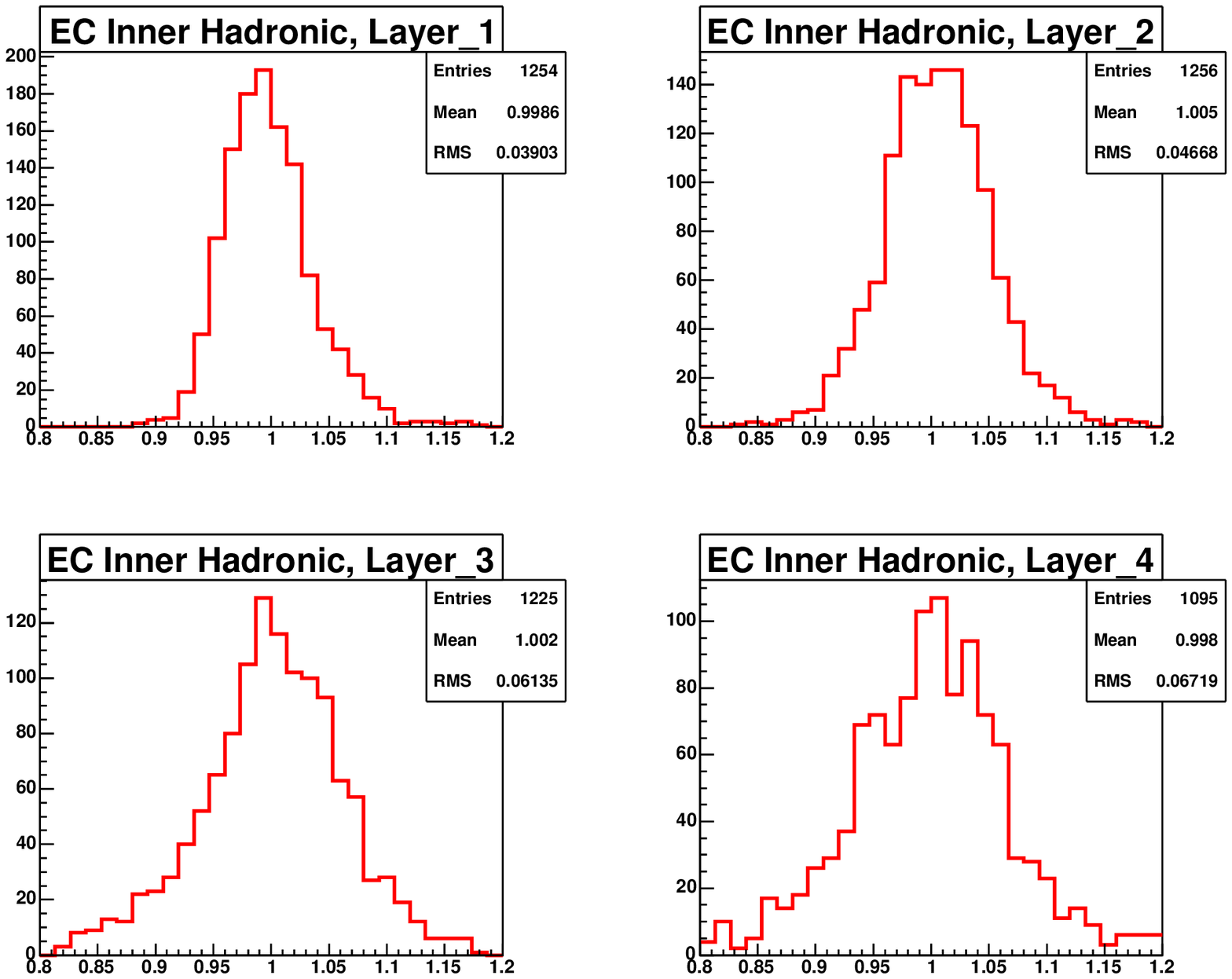}
  \caption{Spread of calibration constants for the first hadronic layer of the calorimeter. The constants are separately plotted for the central, EC inner-hadronic part.}
\label{Proj}
\end{figure}

With the calibration method described, calibration constants on cell
level have been determined for the whole $\eta$ region with available
trigger information (up to $|\eta|$ of 3.2). Due to statistical
limitations in our calibration data sample, for the inter cryostat
region and for the region of $|\eta|$ above 2.4 a calibration on tower
level is used only.

In Fig. \ref{Proj}, as an example, the spread of the calibration
constants is plotted for the first hadronic layers of the detector
separated into two regions: central calorimeter and EC inner hadronic
calorimeter. The calibration constants are mainly in the range of
0.90-1.15 and the root mean squares are at the order of
0.05. Calibration constants are slightly smaller for the central region
compared to the EC and for the first hadronic layer compared to the
other hadronic layers.  The plots of the spread of constants have
tails resulting from outliers with higher constants. For the central
calorimeter this is mainly due to a single module, whose contribution
is plotted separately with dotted lines.  In the EC inner hadronic part
this outliers are from the region of higher $\eta$'s with the lack of
statistics. In these regions only a calibration on tower level is
aimed anyway.

The error estimation was done with a MC method: we generate toy
simulations of the data with known miscalibrations and compare to the
fitted calibration constants of our calibration procedure. The central
calorimeter is now calibrated with the precision of the order of 1\%, for
the high $\eta$-regions it is a few per cent. 

\begin{figure}[p]%\hspace{-1.5cm}
\includegraphics[width=.9\textwidth,height=.28\textheight,bbllx=0pt,bblly=570pt,bburx=369pt,bbury=737pt,clip=,]{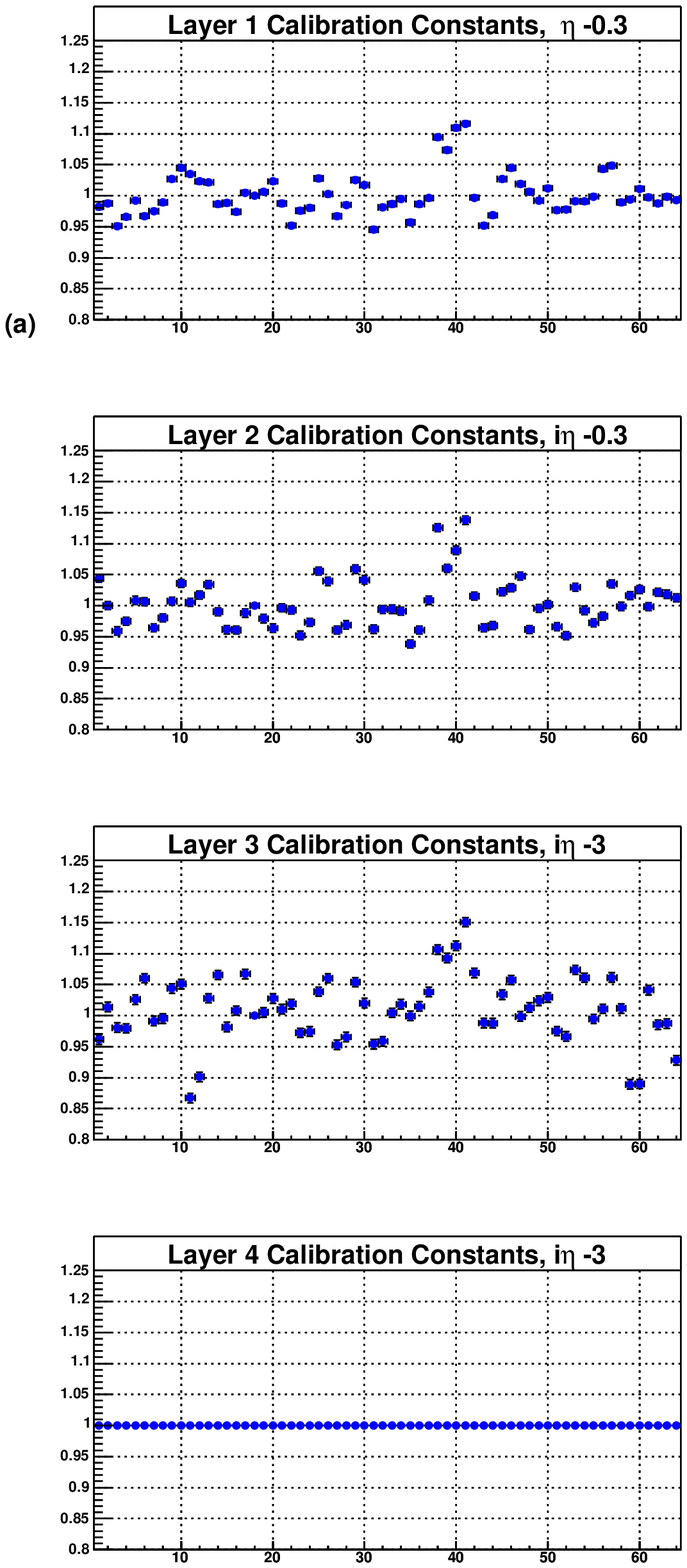}
%\caption{Calibration constants for the first FH layers of the i$\eta$ ring -3.}  
%\label{L3} 
\end{figure}

\begin{figure}[p]%\hspace{-1.5cm}
\includegraphics[width=.9\textwidth,height=.28\textheight,bbllx=0pt,bblly=570pt,bburx=369pt,bbury=737pt,clip=,]
{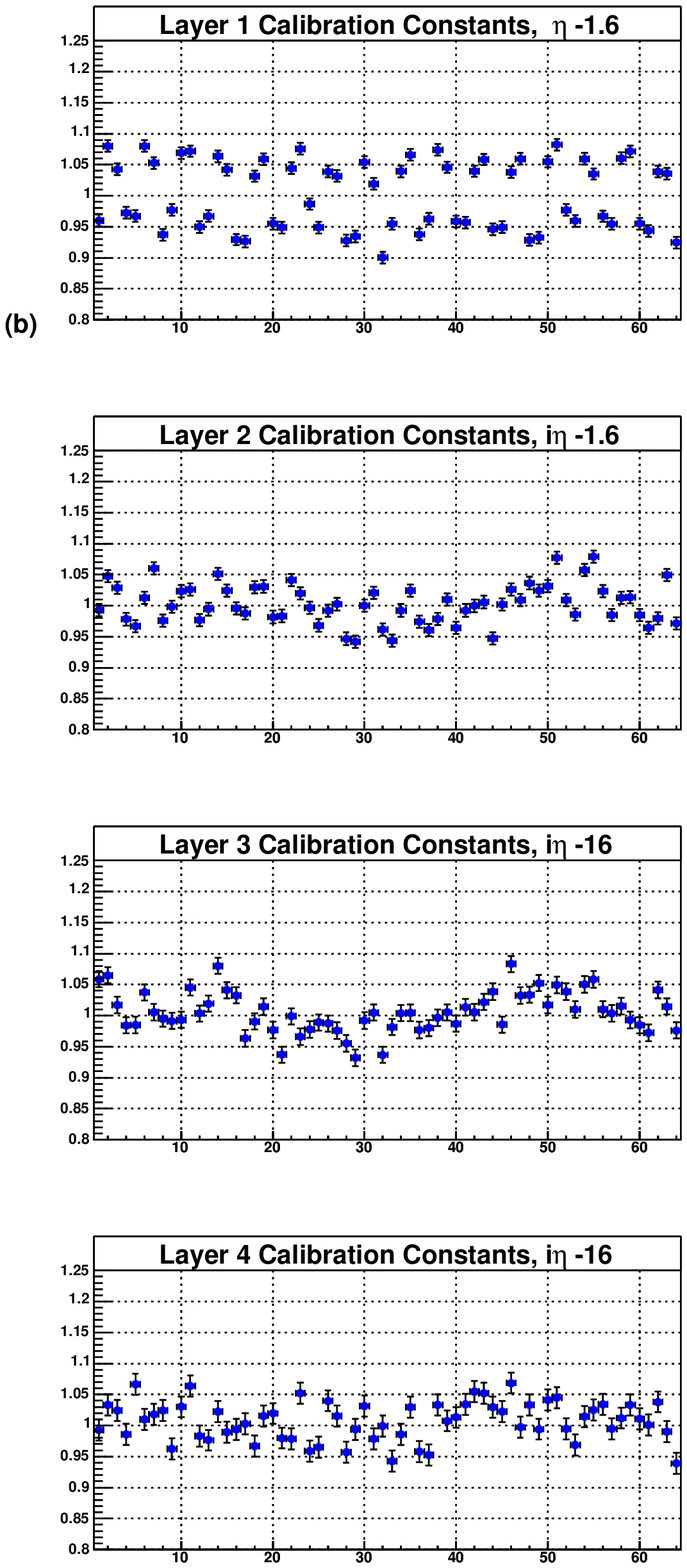}
%\caption{Calibration constants for the first FH layers of the i$\eta$ ring -16.}  
%\label{L-16} 
\end{figure}

\begin{figure}[p]%\hspace{-1.5cm}
\includegraphics[width=0.9\textwidth,height=.28\textheight,bbllx=0pt,bblly=570pt,bburx=369pt,bbury=737pt,clip=,]{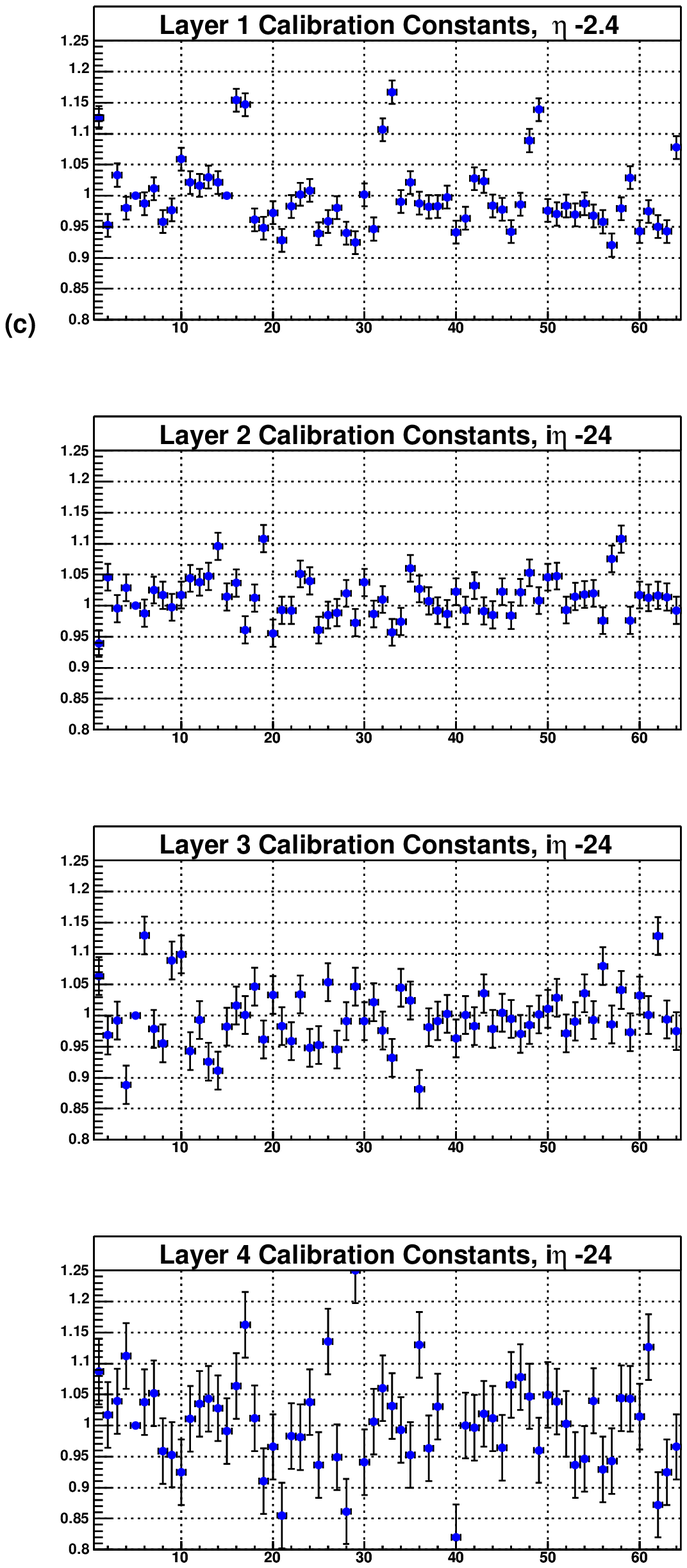}
\caption{$\phi$-intercalibration constants for the first hadronic layer at three different $\eta$-rings at $\eta$ -0.3, -1.6 and -2.4 respectively. The 64 constants corespond to the 64 calorimeter segments in $\phi$-direction.}  \label{L} \end{figure} 

In general the energy response of the modules is less uniform than it
was in Run I. The dominant reason for this is the short integration
time in Run II. This amplifies the effect of the finite precision of
the calorimeter modules. The electron drift time across the 2.3 mm LAr
gap is at the order of 450 ns. While in Run I the integration time was
essentially ``infinite'' on the time scale of this drift time, with
the shorter Run II integration time we cut into the signal. The
$\phi$-intercalibration accounts for these charge collection effects. 

This is illustrated in three examples, where three extreme cases have
been choosen. One is a whole module out of the 16 modules in the
central calorimeter which has a low response, thus had to be boosted
up.  As an example we present Fig.~\ref{L}(a), the calibration constants
at $\eta$ of -0.3, but the same pattern is visible in the whole fine
hadronic central calorimeter. The spread of calibration constants due
to this module is plotted in Fig.\ref{Proj}, (dashed line).

In addition, the effect at the edges of this module is stronger and
such cells need to be boosted more than at the center of the module. This
kind of charge collection effects do not only concern this particular
module, they are also visible throughout the hadronic calorimeter. At
a closer look to Fig.~\ref{L}(a), a similar pattern can be recognized
for all the modules included in the discussed plot. The same
inefficient charge collection is clearly visible in some
cells of the end cap calorimeter which are at the boundaries of this
calorimeter section. The effects are strongly enhanced by the
additional calorimeter borders. An example is plotted in
Fig.~\ref{L}(b) for the $\eta$-ring at -1.6.

The inner hadronic calorimeter was built on one module, thus charge
collection effects in the inner hadronic part are of different kind
compared to the central part. The inner hadronic modules and
absorption layers have been mounted together from half-circles. These
modules and absorption layers are oriented within $\pm 90^{\rm o}$
with respect to each other to obtain a structure without any
gaps. Charge collection effects due to these rotated semi-circles are
visible in the calibration constants. An example is plotted in
Fig.~\ref{L}(c) for the $\eta$-ring at -2.4. Four groups consisting of
two constants need to be boosted up due to charge collection effects.

\section{$\eta$-intercalibration  of the \D0 \ HCAL}

The next step in our calibration procedure is the
$\eta$-intercalibration, where we determine overall calibration
factors for each $\eta$-ring. These will be 64 constants for
$-3.2<\eta < 3.2$ on top of the $\phi$-intercalibration constants. At
this stage the EM layers of the calorimeter are already calibrated and
the hadronic cells are equalized in $\phi$. Our aim at the
$\eta$-intercalibration is to determine a relative weight between the
EM and the hadronic calorimeter which yields the best jet energy
resolution. The necessary consequence of this procedure is well known
\cite{Wigmans:2000vf}, the jet response will be non-linear. 
This is due to the fact that the sampling fraction decreases
considerably as the shower develops, because the calorimeter response
is smaller for the soft $\gamma$ component in the tail of the shower
than for mips.  The fraction of energy deposited in the hadronic part
of the calorimeter will rise with the energy. The shower starts later
and the sampling fraction will rise in this section for the above
reasoning. Since a high sampling fraction gives less fluctuations, the
hadronic part of the calorimeter demands a higher weight for the best
energy resolution. Consequently there are no single optimal constants
for all energies. The default constants have been chosen to be
optimal for jet of 45 GeV which satisfies the vast majority of physics
program at the \D0 \ detector.

\begin{figure}[t]\hspace{-1.5cm}
\includegraphics[width=.9\textwidth]{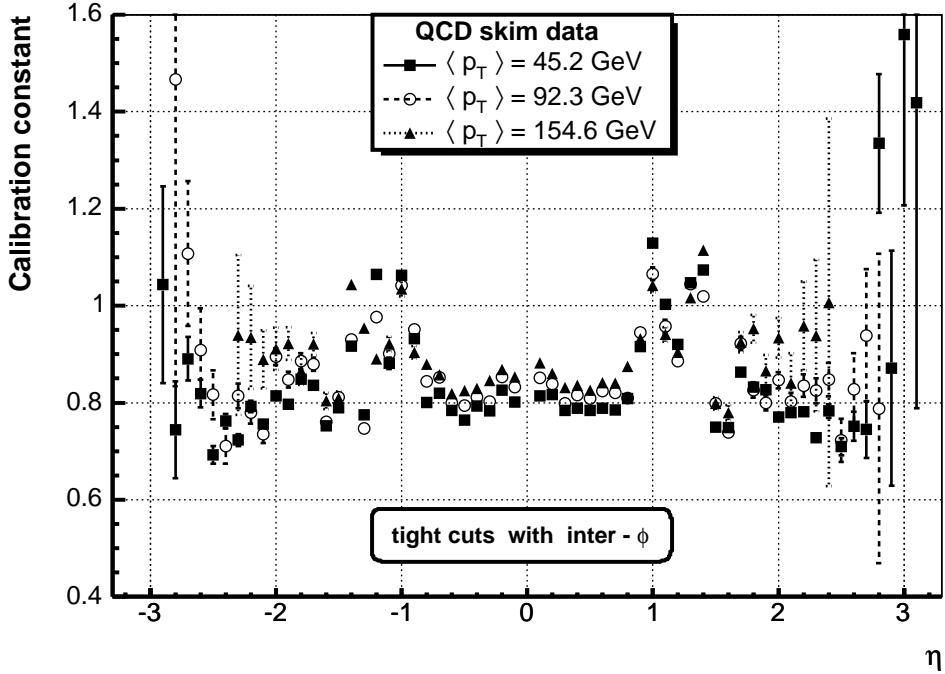}
\caption{$\eta$-intercalibration constants for three different mean jet $p_T$ values of 45.2 GeV, 92.3 GeV and 154.6 GeV respectively. Constants are on top of the older ones which roughly reproduced the right sampling fractions.}  
\label{eta} 
\end{figure}

This discussed non-linearity does not imply that only a certain jet
energy is measured correctly.  With the presence of the amount of dead
material in front of the calorimeter it is anyway non-linear
regardless of the weights. The non-linear calorimeter response to jets
is corrected with an energy-dependent Jet Energy Scale \cite{jiri}.

For the $\eta$-intercalibration we used a sample of QCD dijet events
where the total missing $p_T$-fraction of the events was minimized by
weighting the hadronic calorimeter cells within the jets. Only well
reconstructed back-to-back two jet events have been selected and an
average jet $p_T$ well above the trigger threshold was required.

Results of the $\eta$-intercalibration are plotted in Fig.\ref{eta}
for three different mean jet $p_T$'s. As discussed, calibration
constants rise with the jet $p_T$, however there was no appreciable
dependence on the jet cone size. These are constants that are on top
of the older ones which roughly reproduced the right sampling
fractions. There is a discontinuity visible in the constants which is
due to the fact that there are no EM cells for $1.2 < |\eta | < 1.4$.
The large error bars at the high $|\eta|$ regions is due to a limited
statistics. The correction factor for the regions of $2.0 < |\eta| <
2.7$ are however stable, thus we choose to extrapolate the mean
values of this range to higher $\eta$-values rather than to use the
constants with the large errors.

\begin{figure}[t]%\hspace{-1.5cm}
\includegraphics[width=.9\textwidth]{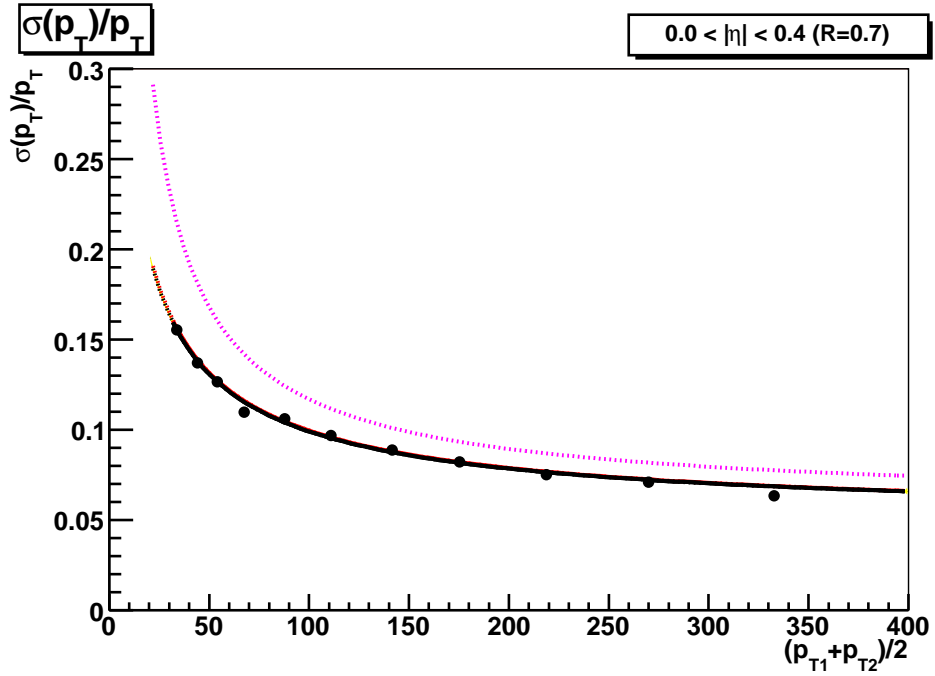}
\caption{Jet $p_T$-resolution in dijet events from the earliest Run II result (dotted line) and 
with the fully calibrated 1 fb$^{-1}$ data set (solid line) in the
$\eta$-range of $0.0 < |\eta | < 0.4$.}  \label{resolution}
\end{figure}

After the full hadronic calorimeter calibration the jet
$p_T$-resolution were re-determined using dijets and the same 1
fb$^{-1}$ sample to account for the improvements due to the hadronic
calibration and Jet Energy Scale. The result is plotted in
Fig.\ref{resolution} for the $\eta$-range of $0.0 < |\eta | <
0.4$. The dotted line is the earliest Run II result and the solid line is
with the fully calibrated 1 fb$^{-1}$ data set. The hadronic
calibration lead to significant improvements in the central region
(ca. 15\% improvement at the energy range of Higgs and top decays).

\end{document}